\documentclass[9pt,conference]{IEEEtran}
\usepackage{amssymb,amsthm,amsmath,array}
\usepackage{graphicx}
\usepackage[caption=false,font=footnotesize]{subfig}
\usepackage{xspace}
\usepackage[sort&compress, numbers]{natbib}
\usepackage{stmaryrd}
\usepackage{xcolor}
\usepackage{mathtools}
\usepackage{float}
\usepackage{textcomp}
\usepackage{hyperref}
\usepackage{subcaption}

\newcommand{\R}{\mathbb{R}}
\newcommand{\C}{\mathbb{C}}
\newcommand{\Z}{\mathbb{Z}}
\newcommand{\I}{\mathcal{I}}

\begin{document}
\title{Beyond Exhaustive Sampling: Efficient Rotational Matching via Ball Harmonics}
\author{\IEEEauthorblockN{
        Fabian Kruse\IEEEauthorrefmark{1}, 
        Vinith Kishore \IEEEauthorrefmark{1},  
        Valentin Debarnot\IEEEauthorrefmark{2}\IEEEauthorrefmark{1},   and
        Ivan Dokmani\'c\IEEEauthorrefmark{1}, 
    }
    \IEEEauthorblockA{
        \IEEEauthorrefmark{1} University of Basel, Switzerland\\
        \IEEEauthorrefmark{2} CREATIS, INSA Lyon, Université de Lyon, France.
        }
}
\maketitle
\begin{abstract}
    Cryo-ET allows to generate tomograms of biological samples \emph{in situ}, capturing complex structures in their native context. Despite low signal-to-noise ratio in reconstructed volumes, the large number of copies of the same macromolecules makes it possible to retrieve high-resolution maps by averaging many aligned subtomograms. 
    To keep up with technical advances in the imaging process and the resulting huge amounts of data available, there is a need for scalable, fast and robust procedures to align subtomograms.
    We propose a subtomogram alignment framework based on the ball harmonics expansion that combines frequency- and gradient-based optimization strategies to avoid exhaustive rotation sampling, enabling a speed-up of an order of magnitude compared to current approaches. 
\end{abstract}

\section{Introduction}
Cryogenic electron tomography (cryo-ET) is a 3D imaging technique that allows to study macromolecular complexes in their natural biological setting.
A vitrified sample is rotated on a single tilt axis and penetrated by electrons, while highly responsive sensors capture projections of the sample. Subsequent merging of projections, known as backprojection, computationally results in a 3D reconstruction of the imaged sample, a tomogram.

The intrinsic dose sensitivity of vitrified biological samples results in a large amount of noise, which confines the resolution of the tomogram and hence of individual complexes. Obtaining high resolution maps of particles of interest is achieved by exploiting multiple copies of the same particle within the tomogram. Aligned subtomograms are averaged to recover high resolution features buried in noise \cite{forster_structure_2007}, \cite{forster_subtomogram_2022}, a process called \emph{subtomogram averaging}.

Any molecular interpretation of reconstructed tomograms requires particles to be detected within the tomogram first. Extracted subtomograms then need to be aligned through shifts and rotations. This is a very challenging task as the reconstruction inherently suffers from low signal-to-noise ratio (SNR) due to a limited electron dose in the imaging process. This is exacerbate by the fact that cells are typically very crowded, damping the contrast of the sample and hindering the alignment of particles because of overlaps. Additionally, during the imaging process, samples are  tilted in limited range typically between $\pm 60^\circ$ in contrast to the full $\pm 90^\circ$ which would be required for a complete reconstruction.
This results in a 'missing wedge' of data in Fourier domain and is responsible for smeared artifacts in reconstructions, typical for cryo-ET. 

Determining occurrence and the optimal alignment is commonly achieved through template matching, where a known (simulated or experimental) structure is cross-correlated with a subtomogram in different orientations. This is an immense computational challenge as it corresponds to a non-convex 6D optimization problem over $\R^3 \times SO(3)$.
Fast Fourier Transform (FFT) makes exhaustive translational search feasible, but sampling rotations remains a large computational challenge. 
Therefore, subtomogram alignment is typically started from a rough initial estimate and rotations are sampled iteratively in a close proximity to rotations obtained in previous rounds of alignment e.g. \cite{wan2024stopgap}.

Rotational sampling can be more effective if volumes are expanded in the basis of spherical harmonics, which led to range of work coined \emph{fast rotational matching} (FRM) \cite{bartesaghi_classification_2008}, \cite{xu_high-throughput_2012}, \cite{chen_fast_2013}. \emph{Ball harmonics} are the generalization of spherical harmonics from the sphere to the bal. Similar to spherical harmonics, they allow for exact rotations using Wigner-D matrices. Computing the cross-correlation using the ball harmonics expansion allows us to utilize gradient-based optimization methods naturally, as derivatives exist in closed form.
 
We propose an efficient, GPU-accelerated subtomogram alignment protocol based on ball harmonics. In contrast to prior techniques for fast rotational matching based on pure spherical harmonics, we utilize the more expressive ball harmonics and employ a joint gradient-based and frequency marching strategy to maximize alignment scores. Compared to existing techniques, our approach achieves a significant speed improvement, operating an order of magnitude faster.

\section{Methods}
\subsection{Ball Harmonics}
The ball harmonics are eigenfunctions of the Dirichlet Laplacian and can be written in spherical coordinates $(r, \theta, \phi)$ as:
 \begin{equation*}
     \psi_{k,l,m}(r, \theta, \phi) = c_{lk}j_l(\lambda_{lk}r)Y^m_l(\theta, \phi) \chi_{\left[0,1 \right)}(r)
 \end{equation*}
 for $m \in \{-l,\dots, l\}, l\in \Z_{\geq 0}$ and $k \in \Z_{>0}$, where $c_{lk}$ are positive normalization constants, $j_l$ are the spherical Bessel functions of the first kind, $\lambda_{lk}$ is the $k$-th positive root of $j_l$, $Y^m_l(\theta, \phi)$ are the spherical harmonics and $\chi_{\left[0,1 \right)}$ is an indicator function for $\left[ 0,1\right)$. They solve the Dirichlet Laplacian eigenproblem on the unit ball $B=\{x\in \R^3 : {\left\lVert x \right\rVert} < 1\}$. A function $f:[-1,1]^3 \to \C$ which is supported on the unit ball $B$ can be expanded into the ball harmonics if it can be written as: 
\begin{equation*}
    f(x) = \sum_{(k,l,m) \in \I} \hat{f}_{k,l,m} \psi_{k,l,m}(r, \theta, \phi)
\end{equation*}
 for a finite index set $\I \subseteq \{(k,l,m) : m \in \{ -l,\dots, l\}, l \in \Z_{\geq 0}, k \in \Z_{>0} \} $, where expansion coefficients of $f$ are depicted by $\hat{f}$.
 
Ball harmonics have some favorable numerical properties.
By ordering the eigenfunctions with respect to the magnitude of their real eigenvalue, we obtain a basis that is \emph{frequency-ordered}. This can be useful to lowpass-filter volumes expanded in the basis, by setting coefficients corresponding to high frequencies to zero. 
Secondly, rotations of the basis functions can be expressed as closed-form linear combinations of same-degree basis functions by making use of Wigner D-matrices. In fact, rotations applied to the basis coefficients correspond to rotating the volume in the Cartesian domain, but in contrast to real space rotations, which introduce inaccuracies through interpolation, rotations of the ball harmonics basis are exact. Expanding a volume in the basis of the ball harmonics gives us a convenient way to rotate volumes and hence we call the basis \emph{steerable}.

\subsection{Cross-Correlation}

Both template matching (TM) and subtomogram alignment (STA) often utilize cross-correlation as similarity measure to determine occurrence, position and translation of a particle within a volume. 

Given a template $t$ and a subtomogram $f$, we want to optimize the cross-correlation between $t$ and $f$ as in \autoref{CrossCorrelation}, where $\mathbf{R}_g$ is the rotation operator of $g\in SO(3)$.
\begin{equation} \label{CrossCorrelation}
   \arg \max_{(s, g) \in \R^3\times SO(3)} C_{f, t}(s, g) := \int_{\R^3}  t(x) [\mathbf{R}_g f](x+s) dx.
\end{equation}
Using the ball harmonics expansion of $f$ and $t$ and the steerable property, we can simplify this expression to obtain \autoref{CrossCorrelation_BallHarmonics} as an expression for the rotational cross-correlation for a specific shift $s$. We assume that both $f$ and $t$ are bandlimited by $L_{\text{max}}$.
\setlength{\belowdisplayskip}{0pt} \setlength{\belowdisplayshortskip}{0pt}
\setlength{\abovedisplayskip}{0pt} \setlength{\abovedisplayshortskip}{0pt}
\begin{equation} \label{CrossCorrelation_BallHarmonics}
    C_{f, t}^{L_{\text{max}}}(s, g) := \sum_{l = 0}^{L_{\text{max}}} \sum_{m,m'} \xi^s_{l,m,m'} D_{m,m'}^l(g),
\end{equation}
where $\xi^s_{l,m,m'}$ are coefficients depending on $\hat{t}$ and the coefficients of the shifted tomogram $\hat{f}_s$,  $D_{m,m'}^l(g)$ are the Wigner-D matrix elements of degree $l$ and order $m,m'$ of $g\in SO(3)$.

To counteract the effects of the missing wedge to the matching procedure, we measure the correlation only in the common area in Fourier space \cite{FORSTER2008276}. 

\subsection{Optimization Strategy} \label{sec:Optimization_strategy}
Maximizing \autoref{CrossCorrelation_BallHarmonics} yields rotation $g\in SO(3)$ that maximizes the cross-correlation between $t$ and $f$ for shift $s\in \R^3$ and repeating this over a grid of shifts results in the maximum scoring alignment as in \autoref{CrossCorrelation}.

\begin{figure}
    \centering
    \includegraphics[height=86px]{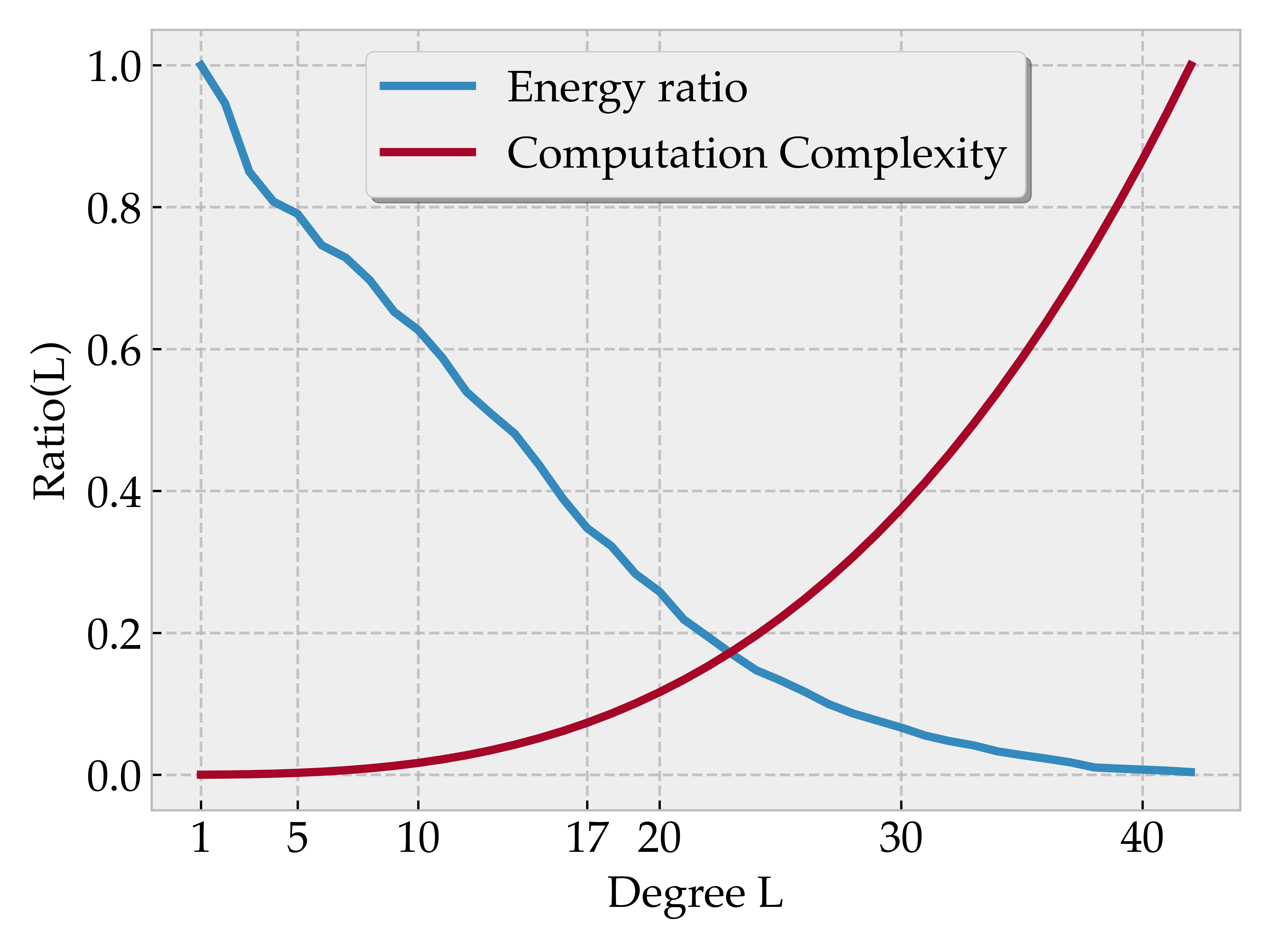}
    \caption{Energy-ratio $ Ratio(L) := \frac{\sum_{l = L+1}^{L_{max}} \mid \xi^s_{l,m,m'}\mid^2}{\sum_{l = 0}^{L_{max}} \mid \xi^s_{l,m,m'}\mid^2}$ and computational effort to evaluate \autoref{CrossCorrelation_BallHarmonics} as a fraction compared to the theoretical computational effort for band $L_{\text{max}}$. Quantities are reported as a function of $L$.}
    \label{fig: energy-ratio}
\end{figure}

The highest accuracy is obtained by exhaustively evaluating \autoref{CrossCorrelation_BallHarmonics} on a fine rotation grid, but huge speed-ups are achievable by truncating at some $L < L_{\text{max}}$. Following a frequency marching strategy (e.g. \cite{barnett_rapid_2017}) by gradually truncated band $L$, we leverage low frequency information that is very efficient to compute.
Restricting the evaluation to only a few bands yields the greatest speed-up and energy-ratio is a powerful heuristic to determine high-energy bands, see \autoref{fig: energy-ratio}.

\begin{figure}
    \centering
    \includegraphics[width=0.95\linewidth]{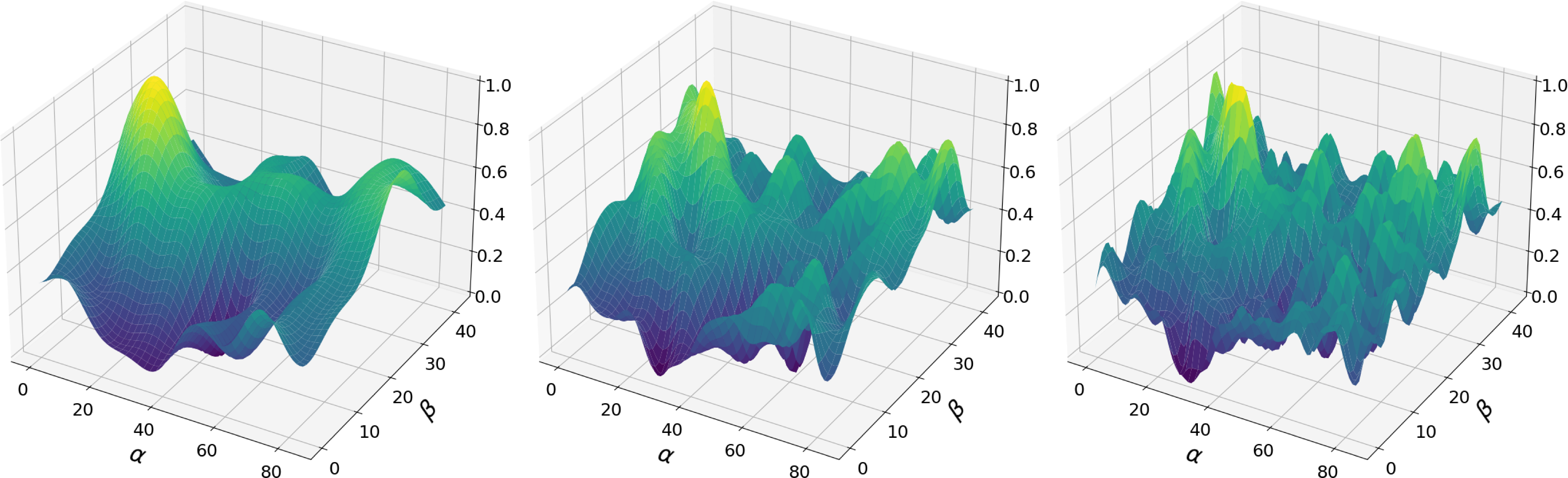}
    \caption{Cross-Correlation of a template and a subtomogram for bands $L_{set}= \{7, 12, 33\}$ as in \autoref{CrossCorrelation_BallHarmonics} for a fixed shift $s$.  For illustration purposes the Cross-Correlation is computed over Euler angles $\alpha$ and $\beta$, while $\gamma$ is fixed.}
    \label{fig:loss-landscapes}
\end{figure}

Since $D^l_{m,m'} \in C^\infty(\R)$ and the derivative exists in closed form \cite{khersonskii_quantum_1988}, we can efficiently compute derivatives of \autoref{CrossCorrelation_BallHarmonics} w.r.t. rotation $g\in SO(3)$.
Additionally, smoothness of the cross-correlation landscape decreases dramatically with increasing bands as is depicted in \autoref{fig:loss-landscapes}. 
Combining frequency marching and gradient-based optimization strategies to maximize \autoref{CrossCorrelation_BallHarmonics}, we obtain the rotation that maximizes the cross-correlation between the two volumes $f$ and $t$ while keeping the time complexity low.
This strategy avoids exhaustive evaluation over a grid of rotations and enables rapid optimization. 

\begin{figure}
    \centering
    \includegraphics[width=0.9\linewidth]{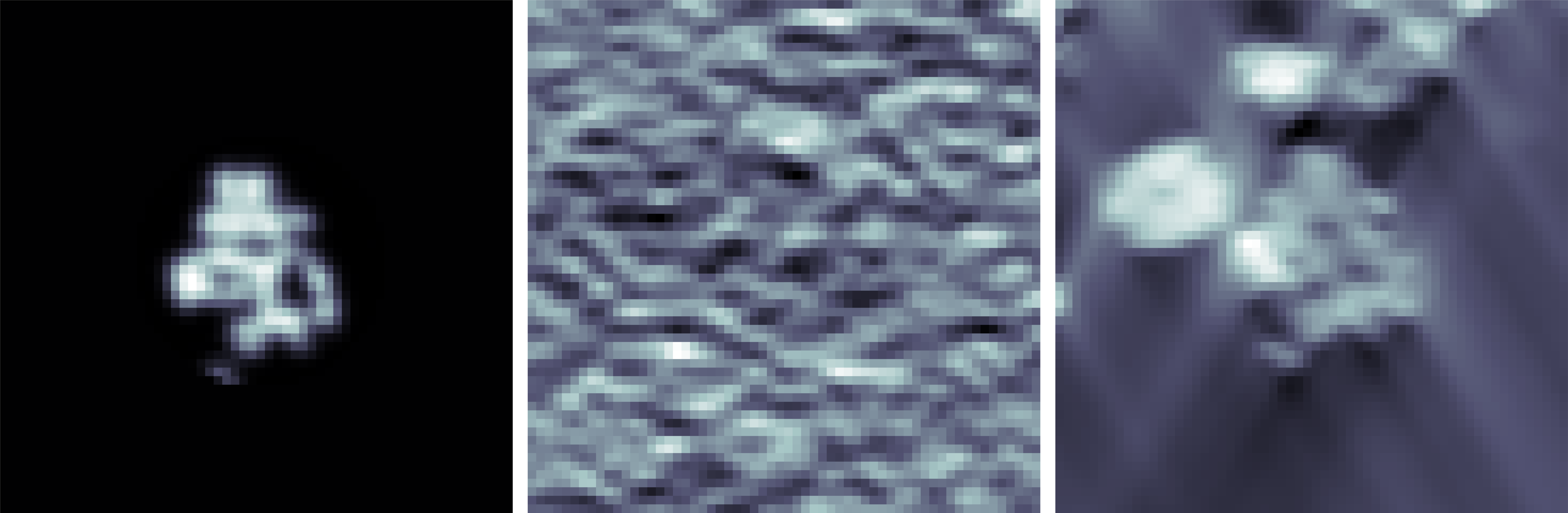}
    \caption{ Central slices through a volume representing: a template, a tomogram and the ground truth map rotated by $g \in SO(3)$ that maximizes \autoref{CrossCorrelation_BallHarmonics}. 
    High similarity between the template and the central particle in the tomogram indicates accurate rotational alignment. Our method recovers the correct rotation with approximately $1.5^\circ$ accuracy. }
    \label{fig:particle_in_subtomogram}
\end{figure}

\section{Experiments and Results}
For experimental evaluation of our method, we focus on the following scenario: given template $t$ and subtomogram $f$, we want to determine shift $s$ and rotation $g$ that maximize the cross-correlation of the two volumes over $\R^3 \times SO(3)$, as given by \autoref{CrossCorrelation}.

Utilizing a recent implementation of the ball harmonics expansion \cite{kileel_fast_2024}, we implement the optimization procedure described in \autoref{sec:Optimization_strategy}. 
The expansion of the volumes is bandlimited to $L_{\text{max}} = 42$, and we selected bands $L_{\text{set}
}= \{7, 12, 33\}$ using the energy ratio, see \autoref{fig: energy-ratio}. Additionally, we use Newton steps to optimize candidate rotations over bands in $L_{\text{set}}$. Initial promising candidates are chosen by local maxima of the lowest selected band $L=7$.

The \emph{Shape Retrieval Competition (SHREC) 2020} \cite{gubins_shrec_2020} 
is based on a set of highly realistic simulated tomograms featuring 10 different tomograms each containing 12 different, known structures. Ground truth maps with the applied rotations are available and used for evaluation.  


Analysis of our method alongside \emph{Fast Rotational Matching (FRM)} as in \cite{chen_fast_2013} reveals that we achieve identical and up to sub-degree accuracy with at least a $10\times$ speed-up. Additionally, our method is successful in crowded tomograms, where FRM often fails e.g. FRM does not correctly match the tomogram in \autoref{fig:particle_in_subtomogram}, while our method achieves $\sim 1.5^\circ$ accuracy.

\section{Conclusion}
We show that subtomogram averaging, a crucial downstream task for high-resolution structural analysis in cryo-ET, can be greatly accelerated by posing cross-correlation as an optimization problem. Leveraging the powerful ball harmonics expansion, we exploit frequency- and gradient-based optimization strategies such as frequency marching and Newton's method. 
Our approach shows very strong results in both matching accuracy and efficiency, which makes it a powerful tool for high-resolution particle reconstruction in cryo-ET.

\newpage
\bibliographystyle{IEEEtran}
\bibliography{references_no_url}
\end{document}